# Phase-change perovskite tunable microlaser

Jingyi Tian[1†], Giorgio Adamo[1†], Hailong Liu[2], Mengfei Wu[2], Maciej Klein[1,3], Jie Deng[2], Norman Soo Seng Ang[2], Ramón Paniagua-Domínguez[2], Hong Liu[2], Arseniy I. Kuznetsov[2], Cesare Soci[1,3]*

[1] Centre for Disruptive Photonic Technologies; TPI, SPMS, Nanyang Technological University, 21 Nanyang Link, Singapore 637371

[2] Institute of Materials Research and Engineering, A*STAR (Agency for Science Technology and Research); 2 Fusionopolis Way, #08-03, Innovis, Singapore 138634

[3] Energy Research Institute @ NTU (ERI@N); Research Techno Plaza, Nanyang Technological University, 50 Nanyang Drive, Singapore 637553

† These authors contributed equally to this work.

*Corresponding author. Email: csoci@ntu.edu.sg

**Abstract:** Since the invention of the laser, adoption of new gain media and device architectures has provided solutions to a variety of applications requiring specific power, size, spectral, spatial, and temporal tunability. Here we introduce a fundamentally new type of tunable semiconductor laser based on a phase-change perovskite metasurface that acts simultaneously as gain medium and optical cavity. As a proof of principle demonstration, we fabricate a subwavelength-thin perovskite metasurface supporting bound states in the continuum (BICs). Upon the perovskite structural phase transitions, both its refractive index and gain vary substantially, inducing fast and broad spectral tunability (1.35 nm/K rate, $\Delta\lambda > 15$ nm in the near-infrared), deterministic spatial mode hopping between polarization vortexes, and hysteretic optical bistability of the microlaser. These features highlight the uniqueness of phase-change perovskite tunable lasers, which may find wide applications in compact and low-cost optical multiplexers, sensors, memories, and LIDARs.

## Introduction

A little more than half a century after the first demonstration of the laser, its varieties and applications have grown enormously. From fundamental spectroscopy, astronomy and quantum optics to atmospheric sensing, optical communications, biomedicine and defense counter measures (*1,2*), many of such applications rely on dynamic tunability of the laser, which typically requires external optical elements and modulation of the resonator/cavity length to control spectral and spatial mode profiles of the beam (*1,3*). In the digital era, where lasers are set to become ever more pervasive and integrated, the use of additional optics poses severe limitations to cost and size reduction. Hence, new types of compact and tunable lasers are in high demand.

Functional metasurfaces relying on phase-change media like chalcogenides (*4-6*), metal-oxides (*7*) and liquid crystals (*8*) offer solutions for dynamically tunable devices in a compact footprint (9). However, while conventional phase-change media are widely exploited in passive reflective and transmissive devices, they are not suitable for active light emitting devices like lasers due to the lack of optical gain. Thus, identifying a class of phase-change materials with tunable refractive index and optical gain could truly transform the landscape of active nanophotonic devices.

Halide perovskites are an emerging class of photonic materials that combine, in a single platform, excellent luminescence properties and high refractive index. This has allowed realization of efficient light emitting devices (*10, 11*) and dielectric nanophotonic structures (*12*) such as high-resolution color displays (*13-15*), photoluminescence enhanced metasurfaces (*16, 17*), and low-threshold lasers (*18, 19*). In addition, due to the strong interplay between inorganic framework and organic ligands, halide perovskites display a rich variety of crystallographic phases which depend on chemical composition (*20*), pressure (*21*) and temperature (*22*). Transitions between these phases are often associated with a large variation of the refractive index and the emission characteristics (*22-24*). However, the potential of halide perovskites as phase-change tunable gain media has yet to be harvested.

In this work, we demonstrate a new type of spectrally and spatially tunable semiconductor microlaser, using a subwavelength-thin halide perovskite film as phase-change tunable gain medium and high-quality factor cavity. The cavity design is based on a dielectric metasurface embossed in the perovskite that supports bound states in the continuum (BICs) (*25-28*), giving a highly directional polarization-vortex emission. The variation in optical constants and luminescence upon phase transition results in a wide spectral tunability over 15 nm range in the near-infrared, with large tuning rate up to 1.35 nm/K - one order of magnitude higher than conventional semiconductor lasers (*29*). At the same time, hopping between two different BICs across the phase transition temperature enables spatial tunability of the laser emission profile between two polarization vortexes carrying opposite topological winding numbers. Furthermore, the thermal modulation of both, spectral and spatial characteristics, shows hysteresis-induced optical bistability. Together, these results demonstrate the potential of phase-change halide perovskites as a new material platform for active nanophotonic devices, such as compact tunable lasers and optical memories.

## Results and discussions

For the demonstration of the phase-change perovskite tunable polarization-vortex microlaser, we select a prototypical halide perovskite, methylammonium lead iodide $CH_3NH_3PbI_3$ ($MAPbI_3$). $MAPbI_3$ has proven to be a dependable high refractive index platform for all-dielectric perovskite metasurfaces and exhibits a tetragonal-orthorhombic structural phase transition at 130-160 K (*22, 23*), associated with considerable modification in refractive index and gain spectrum. The phase-change figure of merit of MAPbI3, $FOM = |\Delta n|/(k_1 + k_2)$=0.78 at 790 nm ($\Delta n$ is the change of refractive index and $k_1 + k_2$ the sum of extinction coefficients of the two phases), is higher than the prototypical phase-change chalcogenide material $Ge_2Sb_2Te_5$ with FOM~0.2 at the same wavelength (*30*) (see Supporting Information I and II).

The metasurface microlaser consists of a square lattice of circular holes patterned in a 250 nm thick MAPbI3 film, sandwiched between a thick polydimethylsiloxane (PDMS) layer and a quartz substrate (*n*=1.5), as depicted in Fig. 1A. Upon temperature change, the MAPbI3 perovskite adopts two distinct crystal structures, i.e., the room-temperature tetragonal phase and the low-temperature orthorhombic phase (Fig. 1B). As indicated by the optical band diagrams in Fig. 1C, the metasurface supports two high-Q modes (solid lines) within the optical gain region above the absorption edge of MAPbI3 (>750 nm) - an out-of-plane (*x-y* plane) magnetic dipole mode (MDz) and a magnetic quadrupole mode (MQ). The two modes give rise to symmetry-protected BICs (MDz-BIC and MQ-BIC) at $k_x$=0, appearing as singularity point at the centre of far-field polarization vortexes. BICs are perfectly confined optical modes, located in the continuum spectrum of free-space radiation, that allow the realization of lasing cavities with ultra-high Q-factors, low lasing thresholds and vortex emission (*18,26-28*). The Q factors of both MDz and MQ modes of the microlaser diverge to infinity near $k_x$= 0, as denoted by the shaded red and blue areas in Fig. 1C. By controlling the ambient temperature, the two BICs can be tuned across the MAPbI3 structural phase transition. As the temperature decreases from 298 to 130 K, the change of refractive index induces a large shift of the spectral position of MDz-BIC (MQ-BIC), from 797 to 782 nm (from 789 to 776 nm) (Fig. 1C).

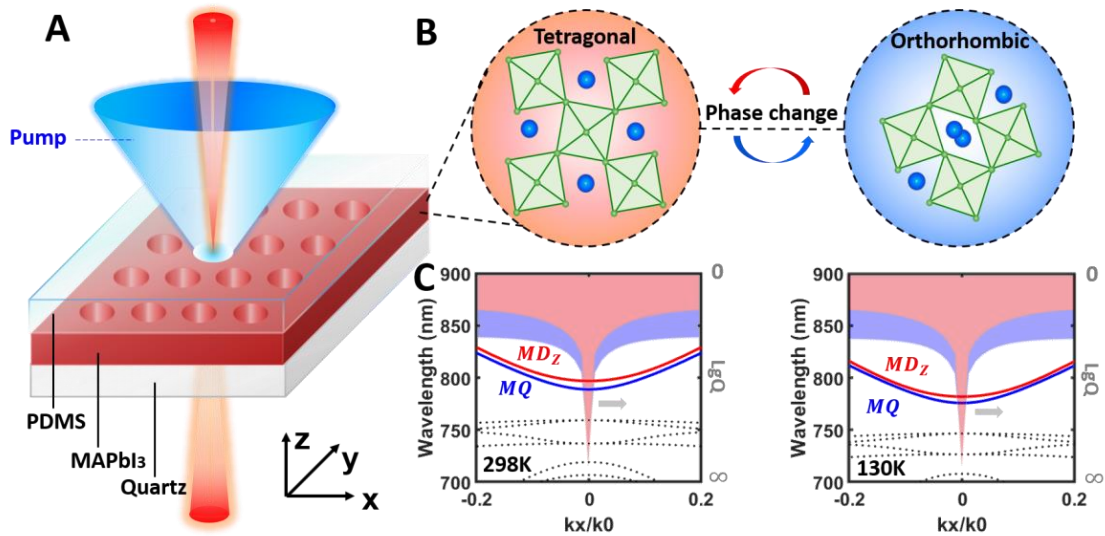

**Fig. 1. Design of a phase-change BIC metasurface (A)** Schematic of a tunable microlaser based on phase-change perovskite BIC metasurface, sandwiched between two n=1.5 layers (quartz and PDMS), pumped with blue laser (λ = 405 nm), and emitting a highly collimated polarization vortex beam in the near-IR. **(B)** Conceptual illustration of the two distinct crystal structures of MAPbI3, the room-temperature tetragonal phase and the low-temperature orthorhombic phase. **(C)** Calculated band structures (solid and dotted lines) of the MAPbI3 metasurface, within the optical gain spectral region, at room temperature, 298 K, (tetragonal phase) and 130 K (orthorhombic phase). The two modes (MDz mode and MQ mode) above the absorption edge are denoted with solid lines. The corresponding Q factors of the two modes are denoted by the shaded red and blue areas, respectively, which diverge at $k_x$=0.

The spectral shift of the BICs with temperature reveals three distinct regions, i.e. a high temperature region with T>160 K where the perovskite is in the tetragonal phase, an intermediate temperature region with 130 K<T<160 K where both tetragonal and orthorhombic phases of the perovskite coexist, and a low temperature region with T<130 K where the perovskite is in the orthorhombic phase (Fig. 2A). When the temperature decreases from 298 to 160 K, the two BICs at 797 and 789 nm gradually redshift by more than 10 nm, corresponding to a tuning rate of 0.1 nm/K. Upon entering the intermediate region (yellow band in Fig. 2A), the peak positions of the MDz-BIC and MQ- BIC blue shift at a much higher rates of -0.97 and -0.85 nm/K, respectively and stabilize at 782 nm and 776 nm when the phase transition is completed. Throughout these regions, the electromagnetic fields of both MDz-BIC and MQ-BIC remain mostly confined within the perovskite film, guaranteeing optimal spatial mode-gain overlap (inset of Fig. 2A).

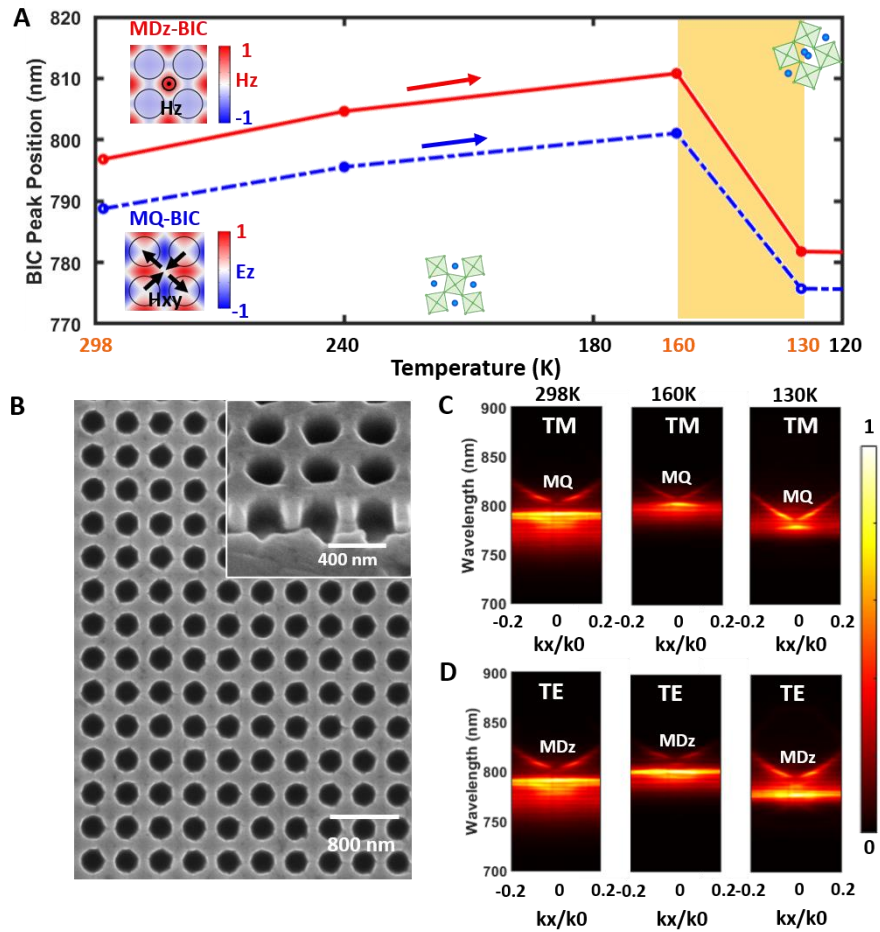

**Fig.2. BIC shifting of the phase-change metasurface upon temperature change. (A)** Calculated shifting of the MDz-BIC (red dashed line) and MQ-BIC (blue dashed line) due to temperature-induced refractive-index change. The inset shows the $H_z$ and $E_z$ field distribution of the two BICs in the *x-y* plane, with antinodes confined within the perovskites film **(B)** SEM image of a MAPbI$_3$ film patterned with a square lattice of circular holes. The scale bar is 800 nm. The inset denotes the cross section of the perovskite metasurface with a scale bar of 400 nm. Measured

**(C)** TM-polarized and **(D)** TE-polarized angle-resolved photoluminescence (PL) at 298K (tetragonal phase), 160 K (tetragonal phase) and 130 K (orthorhombic phase), under 405 nm continuous wave (CW) laser excitation.

The MAPbI$_3$ metasurface is fabricated by nanoimprint lithography (NIL) (see Materials and Methods). The circular holes, with a diameter of 300 nm, are arranged in a square lattice with 400 nm period and imprinted to a depth of ~200 nm within the 250 nm thick film, as shown in Fig. 2B. Besides large area nanostructuring, NIL provides additional advantages such as a dramatic improvement of surface morphology (Fig. S4) and increase of environmental stability of the perovskite film (*31*). Since both BICs are located within the emission region of MAPbI$_3$, its photoluminescence (PL) is expected to efficiently couple to these optical modes with angular emission pattern closely resembling the optical bands in Fig. 1C. Angle-resolved PL is measured at three representative temperatures for transverse magnetic (TM) and transverse electric (TE) polarizations, and the corresponding spectral maps are shown in Figs. 2, C and D. The BIC nature of the MQ mode (TM-polarized) and MDz mode (TE-polarized) is revealed by the PL bands with vanishing emission linewidth near the normal direction ($k_x$=0) while resonances disappear in the absorption region above the MAPbI$_3$ band edge. As predicted by the simulations in Fig. 2A, the PL bands redshift when the temperature decreases from 298 to 160 K, and rapidly blueshift when the temperature decreases further to 130 K.

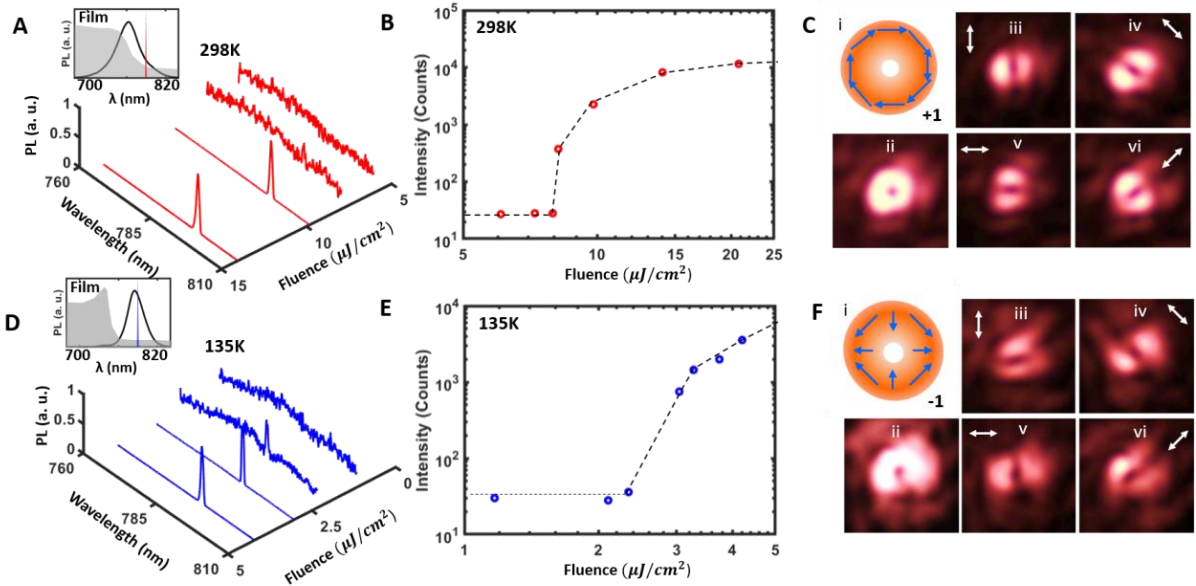

**Fig. 3. Laser performance of polarization-vortex microlasers (A)** Room-temperature emission spectra of the MAPbI$_3$ BIC metasurface as function of pump fluence. The inset shows the absorption coefficient (grey shaded spectrum), PL spectrum (black curve) of a MAPbI$_3$ film, together with the MDz-BIC lasing emission (red shaded peak). **(B)** Light-light curve of the laser at room-temperature. **(C)** i: schematic of a TE-polarized polarization vortex emission with +1 topological winding number; ii: donut-shaped, room temperature, unpolarized far-field radiation pattern of the MAPbI$_3$ BIC metasurface; iii-vi: far-field laser beam profiles passing through a polarizer at four different orientations. **(D)** Emission spectra of the MAPbI$_3$ BIC metasurface as function of pump fluence, at 135K. The inset shows the absorption coefficient (grey shaded spectrum) and PL spectrum (black curve) of a MAPbI$_3$ film, together with the MQ-BIC lasing emission (blue shaded peak). **(E)** Light-light curve of the laser at 135 K. **(F)** i: schematic of a TM-

polarized vector beam with -1 topological winding number; ii: unpolarized far-field radiation pattern of the BIC metasurface, at 135 K; iii-vi: far-field laser beam profiles passing through a polarizer at four different orientations.

Lasing from the BIC metasurface is characterized under frequency-doubled fs-laser pump (λ = 400 nm), with 100 fs pulse duration and 1 kHz repetition rate. The pump-fluence dependence of the room temperature emission spectra are shown in Fig. 3A. A broad emission spectrum is observed at low pump fluences, which is overtaken by a single narrow lasing peak (λ = 797 nm) at higher fluences. The dominant lasing mode is the MDz-BIC, while the MQ-BIC is inhibited by the higher absorption of MAPbI$_3$ at shorter wavelengths (inset of Fig.3A). The onset of lasing at room temperature can be easily identified in the light-light curve in Fig. 3B, yielding a lasing threshold of 8 µJ/cm$^2$. The expected far-field vector distribution of a MDz-BIC is a TE polarization vortex in the *k*-space with the topological winding number +1 (refer to Fig. 3C and discussion in Supporting Information V on the topological nature of the BICs). The measured unpolarized far-field radiation shows the typical donut shape of polarization vortex beams, whereas the polarization dependent beam profiles confirm the winding of the electric field vector around the singularity point (Fig. 3C).

The lasing mode changes from a MDz-BIC (λ = 797 nm) to a MQ-BIC (λ = 791 nm) across the MAPbI$_3$ phase transition (Fig. 3D). As seen in the in the inset of Fig. 3D, the phase transition induces a blue shift of the absorption edge of MAPbI$_3$, resulting in better spectral overlap between the MQ-BIC and the optical gain. Consequently, the lasing threshold reduces to ~2.5 µJ/cm$^2$ at 135 K (Fig. 3E). In this case, the MQ-BIC is expected to generate a TM polarization vortex beam with a -1 topological winding number (Fig. 3F and Supporting Information V). Indeed, far-field imaging of the unpolarized and polarized radiation patterns reveal a polarization vortex underpinned by the MQ-BIC (Fig. 3F).

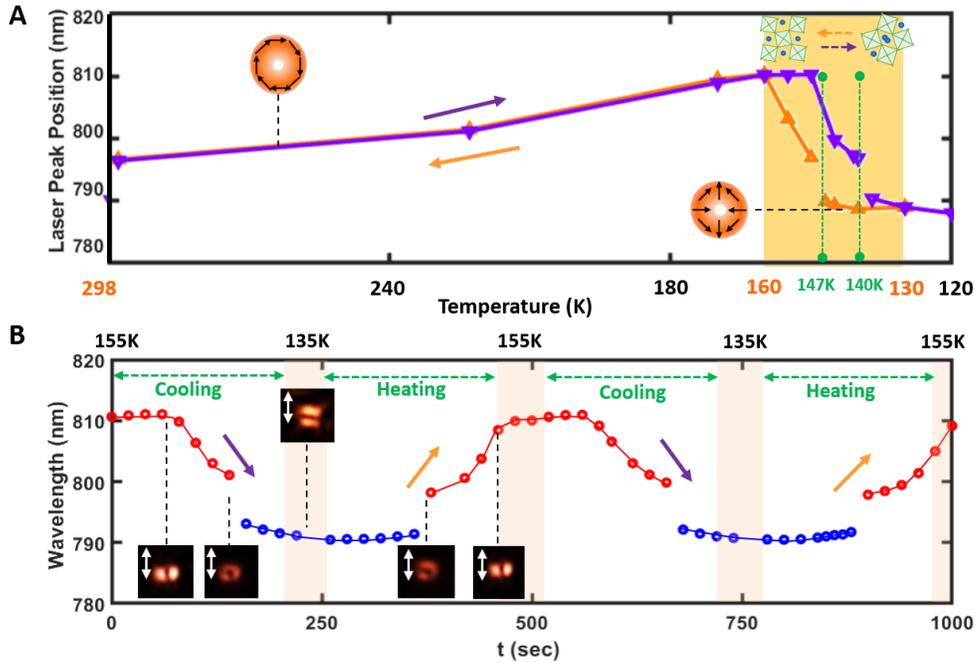

**Fig. 4. Cyclability of the tunable microlaser.** (A) Shifting and switching of the laser modes by decreasing and increasing the temperature. Bistability (hysteresis behavior) can be observed in the phase-transition region (130-160 K). **(B)** Peak position of the bistable laser within two cooling and heating cycles between 135 K and 155 K at the rate of 6 K/min while keeping a constant pump power of 0.64 µW. The two modes are denoted by red and blue colors, respectively. The insets denote their corresponding far-field radiation patterns after a y-polarized linear polarizer at different time period.

Cyclability of the tunable microlaser is studied between 298 and 130 K (Fig. 4 and Fig. S3). Fig. 4A shows the change of laser peak position across a full cooling and heating cycle. The MDz-BIC generated by the perovskite metasurface in the tetragonal phase (160 K< T < 298 K) can be tuned continuously between 797 and 811 nm at a rate of 0.1 nm/K. As the perovskite enters the intermediate region (yellow), the sign of the tuning rate is inverted and the laser wavelength moves between 811 and 788 nm at a much faster rate of -1.35 nm/K. Notably, within the phase transition region, the laser mode switches between the MDz-BIC and the MQ-BIC.

Interestingly, the temperature dependence of the laser emission wavelength does not follow the same path during the cooling and the heating cycles, forming a hysteresis loop. This is underpinned by the hysteresis of the dielectric permittivity of $MAPbI_3$ across the phase transition (*22*). Bistable operation of the microlaser is attainable within the hysteresis region when the laser switches between the MDz-BIC and the MQ-BIC. This allows generating two stable polarization vortex beams with opposite topological winding numbers at the same temperature, that can be addressed within the heating or cooling branches of the temperature hysteresis loop (green-dashed lines at 140 and 147 K). This is further illustrated in Fig. 4B, where the microlaser undergoes several cooling and heating cycles between 135 and 155 K at a rate of 6 K/min. The microlaser can be switched repeatedly between the two stable modes (red and blue curves), showing good retentivity, spectral and spatial mode reproducibility, and minimal intensity degradation over more than 10 cycles (a video of the continuous mode switching is available in Supporting Information). The energy required to switch the microlaser may be further lowered in chemically engineered perovskites with phase transition close-to-room-temperature (e.g., $(C_4H_9NH_3)_2PbI_4$ with phase transition temperature greater than 250 K) (*24*), which may enable fast and efficient dynamic tunability by thermoelectric cooling.

**Conclusions**

Overall, the unprecedented combination of reversible spectral and spatial tunability with optical bistability of the subwavelength-thin phase-change perovskite polarization-vortex microlaser makes it a unique concept for emergent applications of wavelength and space division multiplexing in classical and quantum optical communications, compact optical memory and neuromorphic devices, optical readout sensors and embedded LIDAR systems.

**Acknowledgments:** We are thankful to Guanghui Yuan for useful discussions and acknowledge help of Elena Feltri with optimization of the perovskite films, Bera Kanta Lakshmi with development of the nanoimprint lithography process and Ha Son Tung with angle-resolved photoluminescence measurements.

**Funding:** Research was supported by the Agency for Science, Technology and Research A*STAR-AME programmatic grant on Nanoantenna Spatial Light Modulators for Next-Gen Display Technologies (Grant no. A18A7b0058) , the Singapore Ministry of Education (Grant no. MOE2016-T3-1-006), National Research Foundation of Singapore (Grant no. NRF-NRFI2017-01), and IET A F Harvey Engineering Research Prize 2016.

**Author contributions:** C.S., J.T., G.A., R.P.D. and A.I.K. conceived the idea. J.T. performed the numerical simulations and theoretical analysis. J.T. and G.A developed the setup to test lasing regime and conducted the laser characterization measurements. J.T., G.A. and C.S. wrote the first draft of the manuscript. M.W. performed the angle-resolved photoluminescence measurement. M.K. synthesized the perovskite film. Hailong L., J.D. and N.S.S.A. performed the nanoimprint lithography process under the supervision of Hong L. All authors contributed to the writing of the manuscript.

**Competing interests:** Authors declare that they have no competing interests.

**Data and materials availability:** All data are available in the main text or the supplementary materials.

## Supplementary Materials

Materials and Methods

Supplementary Text

Figs. S1 to S6

Movies S1 to S2

# Supplementary Materials for

## Phase-change perovskite tunable microlaser

Jingyi Tian[1†], Giorgio Adamo[1†], Hailong Liu[2], Mengfei Wu[2], Maciej Klein[1,3], Jie Deng[2], Norman Soo Seng Ang[2], Ramón Paniagua-Domínguez[2], Hong Liu[2], Arseniy I. Kuznetsov[2], Cesare Soci[1,3]*

[1] Centre for Disruptive Photonic Technologies; TPI, SPMS, Nanyang Technological University, 21 Nanyang Link, Singapore 637371

[2] Institute of Materials Research and Engineering, A*STAR (Agency for Science Technology and Research); 2 Fusionopolis Way, #08-03, Innovis, Singapore 138634

[3] Energy Research Institute @ NTU (ERI@N); Research Techno Plaza, Nanyang Technological University, 50 Nanyang Drive, Singapore 637553

† These authors contributed equally to this work.

*Corresponding author. Email: csoci@ntu.edu.sg

**This PDF file includes:**

>Materials and Methods
>Supplementary Text
>Figs. S1 to S6
>Captions for Movies S1 to S2

**Other Supplementary Materials for this manuscript include the following:**

>Movies S1 to S2

**Materials and Methods**

**1. Sample preparation**
**(1) Film preparation**

MAPbI$_3$ thin films are fabricated from 1.2 M precursor solution of CH$_3$NH$_3$I (Dyesol) and PbI$_2$ (99.99%, TCI) (molar ratio 1:1) in anhydrous dimethylformamide (DMF, Sigma-Aldrich). As prepared solution is magnetically stirred overnight at room temperature in N$_2$ filled glovebox, then filtered by a polyvinylidene fluoride (PVDF) syringe filter (0.45 μm) and left on the hot plate at 373 K for one hour before spin-coating. Prior to perovskite deposition, quartz substrates are cleaned by immersing in the following solution: 2 mL of Hellmanex II (Hellma Analytics) in 200 mL of deionized (DI) water at 353 K for 10 min. Subsequently substrates are rinsed with DI water and dried with the flow of nitrogen followed by oxygen plasma cleaning treatment. The perovskite precursor solution is spin-coated onto the quartz substrates with a speed of 4900 rpm for 30 s using antisolvent engineering method. Toluene is drop-casted on the substrates 5 s after the start of the spin-coating program. The resulting films are finally annealed at 373 K for 15 min.

**(2) Mold Fabrication and Thermal Nanoimprint Lithography Process**

*Master mold fabrication*. A thickness of 170 nm HSQ (Hydrogen silsesquioxane, XR-1541-006) is spin-coated on Si substrate at 1500 rpm for 60 sec. The designed patterns are exposed with an e-beam lithography system (ELS-7000, Elionix) at an acceleration voltage of 100 kV, current of 500 pA and dose of 7600 μC/cm$^2$. The exposed sample is developed in salty developer (KOH 1%, NaCl 4%, DI water 95%) for 60s, and rinsed with DI water for 60s. The Si is etched by ICP (Inductively Coupling Plasma, Oxford plasmalab 100 ICP/DRIE) with mixture of HBr (50 sccm) and O$_2$ (3 sccm) gases at 5 mTorr, CP power of 800 W and RIE power of 150 W. After Si etching, the HSQ mask is removed by soaking the sample in buffered hydrofluoric acid for 5 minutes and then rinsed with DI water.

*Thermal Nanoimprinting*. The master mold is coated with silane as anti-sticking layer. Thermal nanoimprinting process is performed with Obducat NIL-60-SS-UV-Nano-Imprinter at 30 bar and 90°C. The master mold is putting on top of the perovskite film, covered with plastic sheet as protection layer. The imprinting time is 30 min, and then the sample is cooled down to 30°C before demoulding process. The printed sample is manually demoulded from the master mold.

**2. Numerical Simulation**

The optical bands and quality factor of perovskite metasurfaces are calculated using three-dimensional finite-element method (COMSOL). One single unit cell, consisting of a circular hole at the center, is simulated when embedded in a homogeneous background (n = 1.5). Periodic boundary conditions are adopted in both x and y directions and perfectly matched layers (PML) along the z-direction are constructed. The optical bands and the corresponding far-field vector distributions in the momentum space are calculated by eigen-frequency solver by sweeping in-plane wavevectors.

**3. Optical characterization**

**(1) Angle-resolved photoluminescence map**

The angle-resolved photoluminescence (PL) spectra are measured by back focal plane (BFP) spectroscopy. The sample is placed in a cryostat (Janis Research ST-500), cooled by liquid nitrogen and connected to a temperature controller (Scientific Instruments Model 9700). The sample is positioned at the image plane of an inverted optical microscope (Nikon Ti-U) with a long working distance objective (50X, NA 0.55). The pump beam from a continuous-wave laser at $l = 405$ nm is expanded by a lens before exciting the sample through the 50X objective, giving a pump spot diameter of ~ 30 mm. The PL signal from the sample is collected through the same 50X objective and a long-pass filter, followed by a series of lenses that image the back focal plane of the objective onto the entrance slit of a spectrometer (Andor SR-303i). The slit is 100 mm in width and is aligned with the x axis of the sample, i.e. collecting the emission in the x-z plane. A grating with 150 lines/mm, blazed at 800 nm disperses the light past the slit, resulting in an angle- and wavelength-resolved PL map, eventually captured on a 2D EMCCD camera (Andor Newton 971).

**(2) Laser characterization**

Photoluminescence of the metasurface is measured by a frequency doubled Ti:Sapphire laser (400 nm, using a BBO crystal) from a regenerative amplifier (repetition rate 1 kHz, pulse width 100 fs, seeded by Mai Tai, Spectra Physics). The pumping laser is focused by a convex lens (with focus length of 3 cm) onto the top surface of the sample and the spot size on the sample is about 50 µm. Emitted light and corresponding fluorescence microscopy image are collected on the backside of the metasuface by a 5X objective lens coupled with a CCD coupled spectrometer (Acton IsoPlane SCT 320) and a camera, respectively. An attenuator and an energy meter are used to tune and measure the pumping density. For laser characterization under different temperatures, the metasurface is mounted inside a Linkam Stage with temperature controller. The temperature is changed by 5 K/min and stabilizes at each target temperature for 5 mins before conducting measurement.

**Supplementary Text**

**I OPTICAL CONSTANTS OF MAPBI3 AT DIFFERENT TEMPERATURES**

The optical properties of MAPbI$_3$ within the gain spectral region (*22*) are shown in Fig. S1. As shown in Fig. S1 (a), it exhibits relatively high refractive index for both orthorhombic and tetragonal phases, covering a broad temperature range and allowing to effectively confine and modulate light at subwavelength scale. The structural phase transition induces significant tunability of the optical constants with $\Delta n \sim 0.2$ between room-temperature tetragonal phase and 130K orthorhombic phase near the wavelength of 785 nm.

In Fig. S1 (b), the absorption edge of MAPbI$_3$ in tetragonal phase is at about 770 nm at room temperature and continuously redshifts to about 790 nm near 160 K before entering orthorhombic phase. Through phase transition, the absorption edge dramatically blue shifts to about 740 nm at 130 K.

In Fig. S1 (c), the phase-change figure of merit of MAPbI$_3$, $FOM = |\Delta n|/(k_1 + k_2)$=0.78 at 790 nm is higher than conventional phase-change chalcogenide material Ge$_2$Sb$_2$Te$_5$ with FOM ~ 0.2 at the same wavelength (*30*). Here $\Delta n$ is the change of refractive index and $k_1 + k_2$ the sum of extinction coefficients of the two phases.

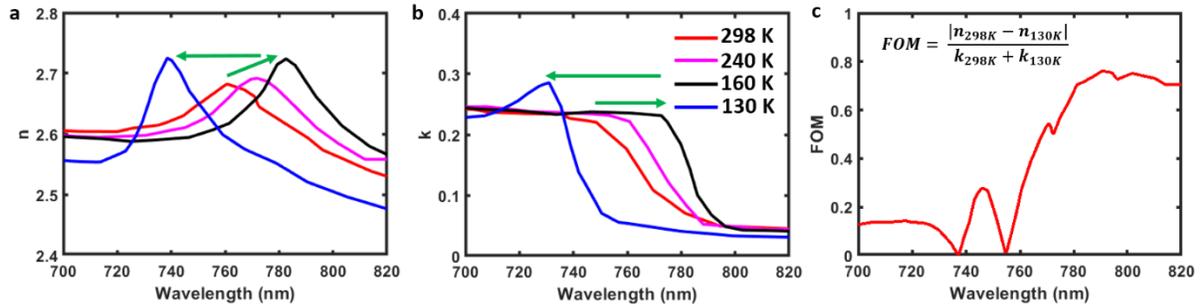

**Fig. S1.** (a) Real part and (b) imaginary part of refractive index of MAPbI$_3$ at different temperatures (*22*) (c) Phase-change figure of merit of MAPbI$_3$.

## II PL FROM MAPbI$_3$ FILM AT DIFFERENT TEMPERATURES

In Fig. S2a and S2b, the spontaneous PL emission from a MAPbI$_3$ film with the thickness of 250 nm under different temperatures are shown as a comparison. At tetragonal phase, the PL peak originating from near-band-edge transition continuously red shifts from 760 nm to 785 nm while temperature decreases from 298 K to 160 K, which is in contrast to most semiconductors. By further decreasing the temperature through the tetragonal-orthorhombic phase, the PL starts to blueshift and a second PL peak at shorter wavelength of 740 nm appears at orthorhombic phase.

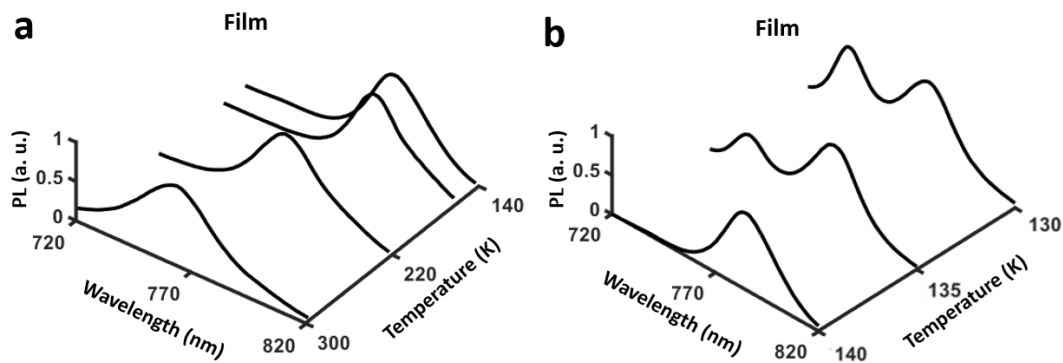

**Fig. S2.** Evolution of PL spectra from the MAPbI3 film (a) from 298 K to 140 K and (b) from 140 K to 130 K.

## III LASER EMISSION FROM THE MAPbI₃ METASURFACE AT DIFFERENT TEMPERATURES

In Figs. S3a and S3b, the measured lasing spectra of the metasurface as a function of decreasing temperature (cooling cycle) are illustrated. The laser peak exhibits a red shift, followed by a rapid blue shift and mode hopping around the phase-transition region.

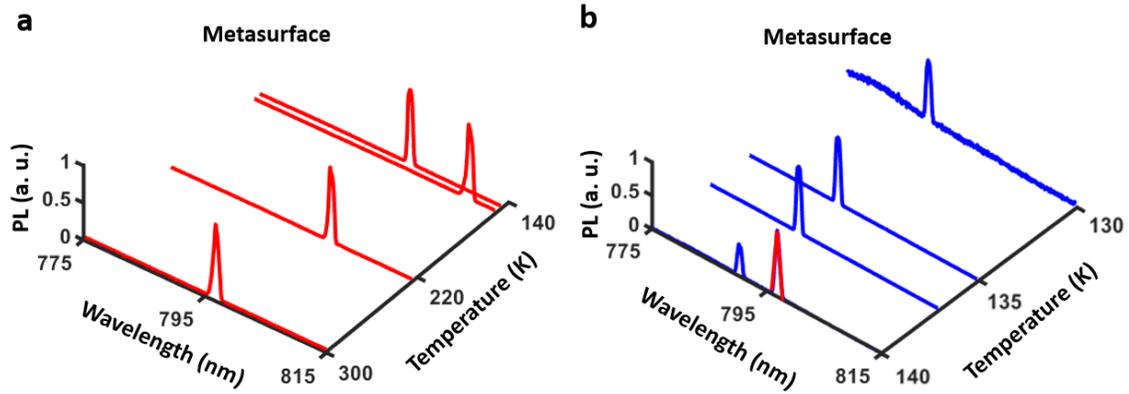

**Fig. S3.** Evolution of laser spectra from the MAPbI₃ metasurface (a) from 298 K to 140 K and (b) from 140 K to 130 K.

**IV PEROVSKITE FILM QUALITY IMROVEMENT AFTER NIL**

Figure S4 illustrates that NIL provides an additional advantage of a dramatic improvement of surface morphology of the perovskite compared with the un-imprinted regions on the same film.

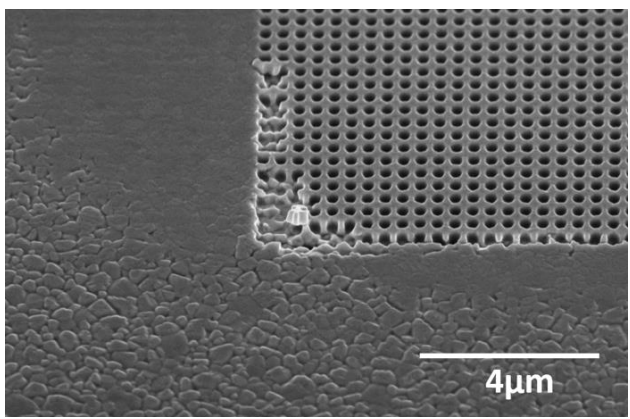

**Fig. S4.** SEM image of the perovskite film after NIL

# V FAR-FIELD POLARIZATION VECTORS OF DIFFERENT MULTIPOLE-BIC MODES

Bound states in the continuum (BIC) correspond to no emission into the far field, which means it is a singularity point of far-field polarization vectors in the momentum space and carries conserved and quantized topological charges as vortex centers. The topological charge ($q$) carried by BIC (i.e., topological winding number) is defined as (26)

$$q = \frac{1}{2\pi}\oint_C dk \cdot \nabla_k \varphi(k), \quad q \in Z \tag{S1}$$

Where $\varphi(k)$ is the angle of polarization vector versus the x-axis and C denotes a closed loop around BIC counterclockwisely in the momentum space. $q$ indicates how many rounds the polarization vector winds around the BIC, which must be an integer. Figure. S3 illustrates schematically how the far-field polarization vectors wind around BICs originating from different multipole modes up to quadrupole modes. Both magnetic dipole (MD) and electric dipole (ED) modes carry topological winding number of +1 while magnetic quadrupole (MQ) and electric quadrupole (EQ) modes carry topological winding number of -1.

Figure S5.b and c shows the calculated far-field polarization vector distributions in the momentum space near the MDz-BIC and MQ-BIC (which are two modes demonstrated in the manuscript) by using the eigen frequency solver in COMSOL. Clearly, they correspond to a topological winding number of +1 and -1, respectively.

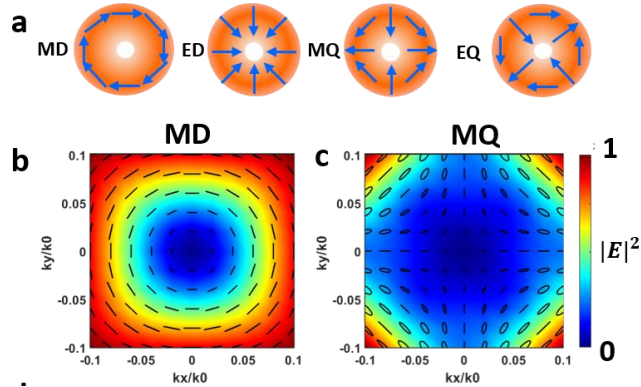

**Fig. S5.** (a) Schematic of far-field polarization vectors of different multipoles in the momentum space. MD, ED, MQ and EQ denote magnetic dipole, electric dipole, magnetic quadrupole and electric quadrupole, respectively. Calculated far-field vector and intensity distribution of the (b) MD and (c) MQ modes of the metasurface in the momentum space, which are illustrated in the manuscript.

## VI LASER EMISSION FROM DIFFERENT SAMPLES AT ROOM TEMPERATURE

Besides the metasurface with the hole diameter of d = 300 nm demonstrated in the manuscript, we also observe that the MAPbI$_3$ metasurface with other hole sizes, e.g., d = 150 nm, d = 200 nm and d = 250 nm can enter the lasing regime with comparable pump fluence and different lasing modes at room temperature. The recorded emission spectra as function of increasing pump fluences are shown in Figure S6a, c, e. The broad emission spectrum, observed at low pump fluences, is modified near the threshold by the appearance of narrow emission peaks, which rapidly overtakes the emission when the pump power exceeds the threshold. When the hole diameter increases, the lasing modes dramatically blue shift. For d = 150 nm, a lasing mode corresponds to EQ mode can be located near the wavelength of 810 nm. When the diameter is increased to 200 nm, EQ mode blue shifts to 804 nm. By further increasing the hole diameter to 250 nm, EQ mode continues to blue shift to about 795 nm while another lower order multipole mode, MD mode, enters the gain regime and start to lase preferentially at about 807 nm. The onset of lasing at room temperature can be easily identified in the light-light curve in Figure. S6b, d, f, where the lasing threshold can be estimated to be around 13 µJ/cm$^2$ for d = 150 nm, 6 µJ/cm$^2$ for d = 200 nm and 8 µJ/cm$^2$ and 10 µJ/cm$^2$ for d = 250 nm, respectively.

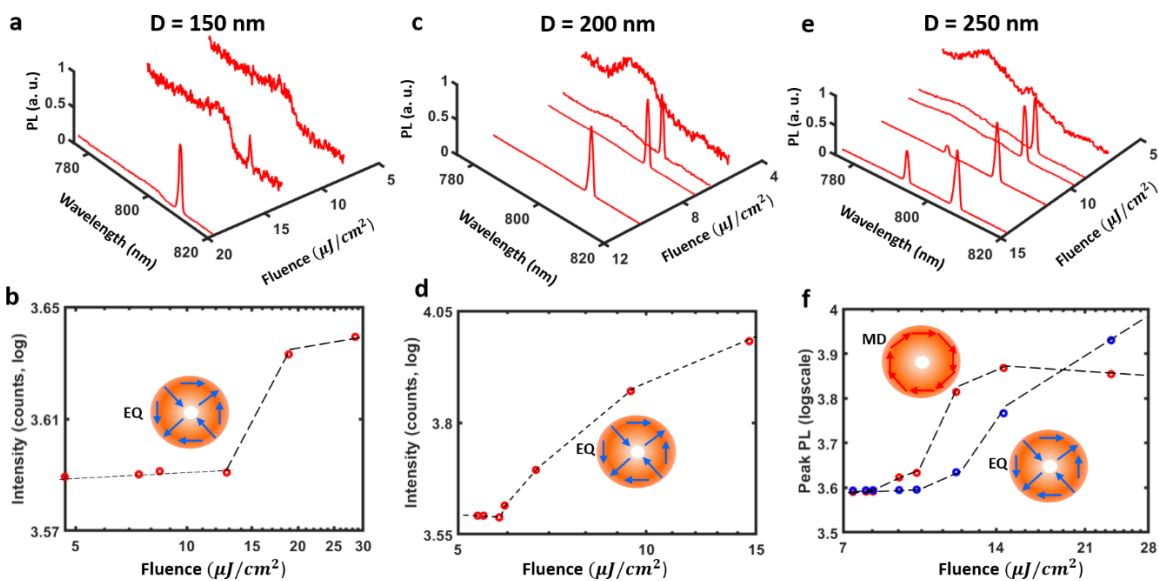

**Fig. S6.** Room-temperature emission spectra of the MAPbI$_3$ BIC metasurface and its Light-light curve with (a), (b) d = 150 nm; (c), (d) d = 200 nm and (e), (f) d = 250 nm as function of pump fluence.

**Movie S1.**

Continuous mode switching during the cooling cycle resolved by a y-polarized linear polarizer.

**Movie S2.**

Continuous mode switching during the heating cycle resolved by a y-polarized linear polarizer.